\documentclass[a4paper,10pt]{article}

\usepackage{graphicx}

\begin{document}


\begin{titlepage}

\begin{center}
\Large{\bf{ Investigating Rare Events by Transition Interface Sampling
}}
\end{center}

\vspace{5mm}
\begin{center}
\large{ Daniele Moroni$^{a,*}$, Titus S. van Erp$^b$, Peter
G. Bolhuis$^a$ }
\end{center}

\vspace{5mm}
\begin{center}
\normalsize{ $^a$Department of Chemical Engineering, Universiteit van
Amsterdam\\ Nieuwe Achtergracht 166, 1018 WV Amsterdam, The
Netherlands }
\end{center}
\begin{center}
\normalsize{ $^b$Laboratoire de Physique \\ Centre Europ\'een de
Calcul Atomique et Mol\'eculaire, \\ Ecole Normale Sup\'erieure de
Lyon, \\ 46 all\'ee d'Italie, 69364 Lyon Cedex 07, France }
\end{center}

\vspace{5mm}
\begin{center}
\large{\bf {Abstract}}
\end{center}
\vspace{2mm}
We briefly review simulation schemes for the investigation of rare transitions
and we resume the recently introduced Transition Interface Sampling, a method
in which the computation of rate constants is recast into the computation of
fluxes through interfaces dividing the reactant and product state.
\vspace{5mm}

\noindent{\it {\bf keywords}: Rare events; Rate constants; Transition
Path Sampling.}

\vspace{5mm}
\noindent\small{$^*$Corresponding author: Tel: +31-20-525-6917; fax:
+31-20-525-5604;\\ email: moroni@science.uva.nl}\\

\end{titlepage}


\normalsize \setlength{\baselineskip}{20pt}


\section{Rare events}

In many complex systems of physical importance transitions take place between
stable states separated by a high (free) energy barrier. Examples are
isomerizations in clusters, chemical reactions, protein folding and crystal
nucleation. Molecular simulations techniques such as molecular dynamics (MD) in
principle enable the computation of the reaction rate constants, the search for
transition states and the exploration of reaction mechanisms. But since the
rate constant of the transition depends exponentially on the activation barrier
height, the expectation time of a transition can easily become orders of
magnitude longer than the molecular timescale which is usually measured in
femtoseconds. Hence, when using straightforward MD the study of these {\it rare
events} is far beyond current computer capabilities.

In the past decade, a number of methods has been proposed that aim to tackle
this timescale problem from different points of view. One class of methods
focusses on  escaping the initial state without making assumptions on the final
state. This can be achieved by, for instance, artificially increasing the
frequency of the rare event in a controlled way. The methods of Voter and
collaborators follow this approach: hyperdynamics \cite{hyperD1,hyperD2} aims
at lowering the energy difference between the top of the barrier and the
initial basin, the parallel replica method \cite{hyperparal1} exploits the
power of parallel processing to extend the molecular simulation time, and
temperature-accelerated dynamics \cite{hyperT1,hyperT2} speeds up the event by
raising the temperature. The idea of driving energy into the system to escape
the basin of the energy minimum in which the system is initially prepared is
also at the basis of conformational flooding \cite{flood}, the Laio-Parrinello
method \cite{LP02,ILP03}, and the enhanced sampling of a given reaction
coordinate \cite{Melchionna00}. Another possible route is to coarse-grain the
molecular dynamics on the fly and explore the resulting free-energy
landscape~\cite{hummer}. Several methods are devoted to the exploration of the
full potential energy surface through all its minima and saddle points.
Examples are eigenvector following \cite{Cerjan81,Doye97}, the
activation-relaxation technique of Barkema and Mousseau, \cite{ART96}, the
dimer method of Henkelmann and J\'onsson \cite{HJ99}, the kinetic Monte Carlo
(MC) approach \cite{G78,V86,FW91}, and the
discrete path sampling of Wales \cite{Wales02,Wales03}.

When the initial and final state are known it is possible to generate paths
connecting the two in the form of a discretized chain of states. This is the
basis of a second class, the so-called two point boundary methods. One option
is to find a minimal energy path on the potential energy surface, as in the
Nudged Elastic Band method of J\'onsson and collaborators
\cite{MJS95,HUJ00,HJ00,JMJ98} and in the string method of E et al.
\cite{ERVE02}, or to find a true dynamical path by minimizing a suitably chosen
action \cite{PP01}. Another possibility is to use modified stochastic equations
of motion that guide the system from the initial to the final state
\cite{ZW99}.

To summarize, one can say that each of the above methods is well suited in one
specific subclass of systems, but they also have their specific drawbacks. Some
are inefficient for high dimensional systems or only give structural
information and neglect the dynamics. Others are designed to find only one
transition or make heavily use on assumptions or prior knowledge of the system.
In complex systems at finite temperature, concepts like the minimum energy (or
action) path or the lowest saddle point are not very useful. The reaction is
rather described by a ensemble of paths. Similarly, one cannot speak of a
particular transition state but only of an ensemble of transition states.

An important quantity describing the kinetics of rare events is the rate
constant. The traditional way to calculate rate constants in complex condensed
matter system is by the reactive flux method based on Transition State Theory
\cite{FrenkelSmit, Keck67, Anderson73, Bennet77, DC78}. This method consists of
two stages. First, the free energy is computed as a function of selected degrees
of freedom describing the reaction from the initial to the final state (the
reaction coordinate), for instance by biased sampling
techniques~\cite{TV74,CCH89,Ciccotti91}. This step is complemented with the
calculation of a dynamical transmission coefficient, by starting short
trajectories from the maximum of the free energy barrier. However, in complex
systems the correct reaction coordinate can be exceedingly difficult to find.
If the reaction coordinate does not capture the molecular mechanism, the biased
sampling methods will suffer from substantial hysteresis when following the
system over the barrier. Moreover, even if the free energy profile is obtained
correctly for this particular (but wrong) reaction coordinate, the corresponding
transmission coefficient will be very low, making an accurate evaluation
problematic.

To overcome these problems, Chandler and collaborators developed the Transition
Path Sampling (TPS) method. This technique gathers a collection of true
dynamical trajectories connecting the states without any \emph{a priori}
assumption of the reaction coordinate. From the ensemble of pathways rate
constants can be calculated and reaction mechanisms can be extracted
\cite{TPS98,TPS98_2,TPS98_3,TPS99}. The method can be combined with parallel
tempering \cite{VS01} and stochastic dynamics can be used for the case of
diffusive barriers~\cite{bolhuis03,CC01}. Successful applications of TPS are,
among others, ion pair dissociation in water \cite{GDC99,MCC00}, alanine
dipeptide in vacuum and in aqueous solution \cite{BDC00}, neutral \cite{RML02}
and protonated \cite{GDC99_2,LRDC01} water clusters, autoionization in water
\cite{GDCHP01}, and the folding of a polypeptide \cite{bolhuisPNAS}. For a
detailed review on TPS see Refs.~\cite{Bolhuis02, Dellago02, BDGC00}. Similar
techniques by Elber and Olender \cite{OE96,ZE00,EMO99} and Doniach et al
\cite{EGJD01} sample discretized stochastic pathways based on the
Onsager-Machlup action. Finally, we mention the topological method of T{\u
a}nase-Nicola and Kurchan \cite{TNK03} in which they suggest to use TPS in
combination with saddle point searching vector walkers.

In this paper we focus on transition path sampling, and in particular on the
Transition Interface Sampling (TIS) method, a recent improvement over the TPS
rate constant calculation \cite{ErpMoBol2003}. We briefly review the TIS method
and give an example of its application. Further details on the derivations,
algorithms and applications can be found in Ref.~\cite{ErpMoBol2003}.

\section{Transition Interface Sampling}

Consider a dynamical system prepared in the initial state $A$. The state is
stable in the sense that trajectories will stay in that state for a time long
compared to the molecular time scale (e.g. vibrations). Eventually, the
trajectories cross the barrier and reach the final state $B$. If the barrier is
sufficiently high the system shows exponential relaxation and the rate constant
$k_{AB}$ is well defined. In what follows we assume that we can compute the
evolution of trajectories in phase space, without further specifying the
details of the dynamics. For instance, we could use deterministic integration
of Hamilton's equations of motion or stochastic dynamics generated by a
Langevin equation.

The starting point of TIS is the partitioning of phase space by interfaces. We
define an order parameter $\lambda(x)$ as a function of the phase space point
$x$ (consisting of positions and momenta of all particles in the system) and
the interfaces $i=0\ldots n$ as the hypersurfaces $\{x|\lambda(x)=\lambda_i\}$.
We assume that the interfaces do not intersect, that $\lambda_{i-1}<\lambda_i$,
and we describe the boundaries of state $A$ and $B$ by $\lambda_0$ and
$\lambda_n$ respectively. In the same spirit of TPS, the TIS method has the
non-trivial advantage that the order parameter $\lambda(x)$ does not have to be
a properly chosen reaction coordinate capturing the essence of the dynamical
mechanism. Instead, it is sufficient that this function is able to characterize
the basins of attraction of the stable states~\cite{Bolhuis02}. The basis of
the TIS method is the microscopic equation for the rate constant
\begin{equation}\label{eq:kab_phi}
k_{AB}= \langle \phi_{n,0} \rangle / \langle h_{\mathcal A} \rangle.
\end{equation}
Here $\langle \phi_{i,0}\rangle$ is the \emph{effective positive} flux from
interface 0 -- i.e. from state $A$ -- through interface $i$, where effective
means that recrossings are not being counted. Equivalently, it is an average
over all phase points that are first crossings through interface $i$ but belong
to trajectories that originated in $A$. The denominator $\langle h_{\mathcal
A}\rangle$ is a normalization factor taking into account all the phase points
for which the corresponding trajectories come directly from $A$ without having
visited $B$ (naturally, this includes the entire region A, and most part of the
basin of attraction of A). In a simulation the ratio in Eq.~(\ref{eq:kab_phi})
is computed by starting a MD simulation in state $A$ and counting the number of
effective crosses with interface $n$, i.e. the number of times it reaches state
$B$, per time unit. Since the transitions are rare, Eq.~(\ref{eq:kab_phi}) is of
no practical use in this form, because the simulation would have to be long
enough to see at least one spontaneous event.  We can improve efficiency by
relating the effective flux $\langle \phi_{i,0}\rangle$ through an interface $i$
to the effective flux $\langle \phi_{i-1,0}\rangle$ through an interface $i-1$
closer to $A$:
\begin{equation}\label{eq:fluxes}
\langle \phi_{i,0} \rangle = \langle \phi_{i-1,0} \rangle \cdot
P_A(\lambda_i|\lambda_{i-1}).
\end{equation}
Here, $P_A(\lambda_i|\lambda_j)$ is the conditional probability that a
trajectory coming from $A$ crosses interface $i$ provided that it has passed
interface $j$. Iteratively substituting Eq.~(\ref{eq:fluxes}) in
Eq.~(\ref{eq:kab_phi}) we can write
\begin{equation}\label{eq:kab_TIS}
k_{AB}=\frac{\langle \phi_{1,0}\rangle}{\langle h_{\mathcal A}
\rangle} \prod_{i=1}^{n-1} P_A(\lambda_{i+1}|\lambda_i)\equiv
\frac{\langle \phi_{1,0}\rangle}{\langle h_{\mathcal A} \rangle}
P_A(\lambda_n|\lambda_1),
\end{equation}
which is the key equation of TIS. Similar equations can be written for the
backward rate constant $k_{BA}$.

We compute the first term in Eq.~(\ref{eq:kab_TIS}) by starting a simulation in
$A$ and counting the number of crossings with interface 1 per unit time. A
statistically accurate value can be obtained by choosing the interface close
enough to the initial state. The second term $P_A(\lambda_n|\lambda_1)$ is
computed using the factorization in Eq.~(\ref{eq:kab_TIS}). First, we generate
an ensemble of paths that start in $A$, cross interface 1 and eventually return
to $A$ or, alternatively, continue to cross the next interface 2. The
probability $P_A(\lambda_2|\lambda_1)$ is the ratio of the number of paths that
reach interface 2 to the total number of sampled paths. Subsequently, we
generate an ensemble of paths starting in $A$ and crossing interface 2, from
which the next term $P_A(\lambda_3|\lambda_2)$ can be obtained, and so on until
we reach interface $n$ and thus state $B$. Each step consists of a sampling of
paths starting in $A$ and crossing interface $i=1\ldots n-1$. This procedure is
similar in spirit to the umbrella sampling technique in which different windows
are employed to obtain free energy profiles \cite{FrenkelSmit}. Note that the
final rate constant $k_{AB}$ is independent of the choice of the interfaces as
long as the first and last one are inside the basin of attraction of the stable
states $A$ and $B$ respectively.

The path ensembles are generated by constructing a random walk in path space
employing a MC algorithm. Fundamental for both the TPS and TIS methods is the
MC move that generates a new path from an existing one, the \emph{shooting
move}~\cite{TPS98_2}. An existing path between the initial and final state is
stored as collection of discrete time-slices. In a shooting move, a time slice
is chosen at random, the momenta of all particles are slightly changed, and a
new trajectory is created by integrating the system forward and backward in
time. The new path is accepted with a probability chosen such that detailed
balance is obeyed. In contrast to TPS, the path length does not have a prefixed
value~\cite{TPS98_2}, but is allowed to vary because the integration of the
equations of motion is stopped when one of the two interfaces of the
corresponding ensemble is reached~\cite{ErpMoBol2003}.

\section{Discussion}

\begin{figure}[hb]
\begin{center}
\includegraphics[width=\textwidth,keepaspectratio]{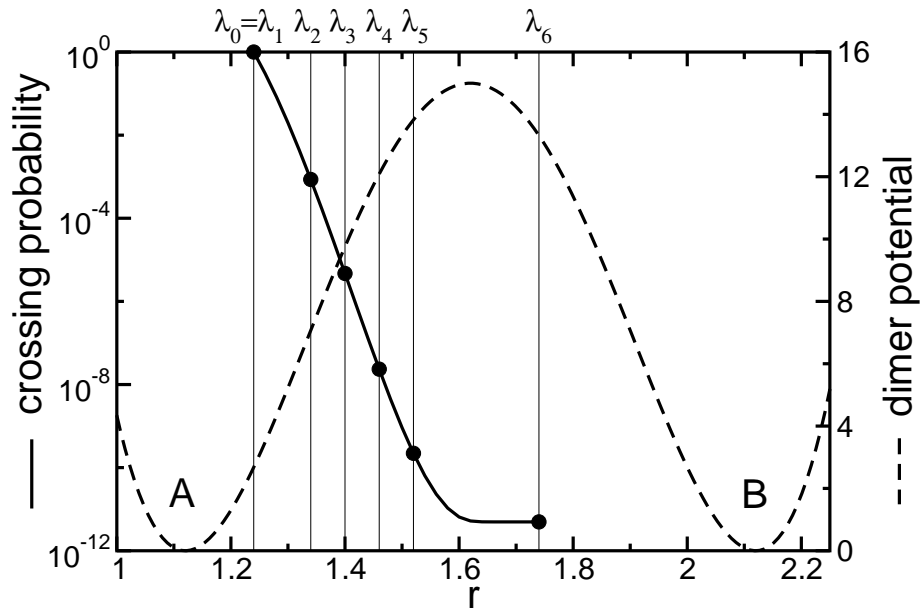}
\end{center}
\caption{
The dimer inter-particle potential (dashed line) and the TIS crossing
probability $P_A(\lambda,\lambda_1)$ (solid line) as function of the order
parameter $\lambda=r$, the dimer inter-particle separation. State $A$ at the
first minimum of the double-well corresponds to a compact state of the molecule
and state $B$ at second minimum to an extended state.  The number of 2D solvent
particles with unit diameter is 25, the density is 0.7, the total energy was
fixed at E=25 (all in Lennard-Jones reduced units~\cite{FrenkelSmit}). We
computed the flux $\langle \phi_{1,0}\rangle / \langle h_{\mathcal A}
\rangle=0.1196\pm 0.0005$, and the crossing probability function
$P_A(\lambda|\lambda_1)$ by matching the partial functions
$P_A(\lambda_{i+1}|\lambda_{i})$, $i=1\ldots 5$ in five windows of path ensemble
simulations.The interfaces in between are indicated by vertical lines. The error
on the data points is within symbol size. The smooth line joining the points was
created using a finer grid of interfaces \cite{ErpMoBol2003}. The horizontal
plateau when approaching state $B$ at $\lambda_{n=6}=1.74$ is an expression of
the commitment of the trajectories to the final stable state. Paths that cross
$r\simeq 1.7$ always reach eventually the final interface without going back to
$A$.  The value of the plateau equals $P_A(\lambda_n,\lambda_1)$.
} \label{fig1}
\end{figure}

We tested the TIS method on the isomerization of a diatomic molecule immersed
in a repulsive fluid, a simple model system used before to test the TPS
algorithms~\cite{TPS99}. We report some details and the results in Fig.
\ref{fig1}; more details can be found in Ref~\cite{ErpMoBol2003}. The forward
rate constant for the isomerization corresponds to an average transition time
$k_{AB}^{-1}=(3.6\pm 0.1)s$ in real units for argon, which is indeed many
orders of magnitude beyond the MD time-step $\sim 4fs$. In Ref.~\cite{ErpMoBol2003}
we showed that TIS is at least a factor 2 more efficient than TPS when
computing the rate constant at equivalent conditions and same final relative
error. The efficiency increases to 5 or more in systems with many recrossings.

The TIS algorithm makes use of the MC moves developed for TPS, and in this
sense TIS can be considered an extension of TPS. TIS retains the good features,
such as the independence of prior knowledge of a reaction coordinate
\cite{Bolhuis02}, and improves on the weaker points, for example, minimizing
computational effort by allowing a variable path length. The spirit behind the
TIS methodology for computing rate constants, however, is different. The
concept of a positive effective flux gives a faster convergence because only
positive terms contribute to the rate. The implementation of the computer
algorithm becomes easier and one can apply multidimensional or discrete
interface order parameters. These advantages make TIS more efficient in terms
of computational effort. Finally, in some cases a better characterization of
true reactive paths can be achieved and non-true recrossings can be avoided
through a proper choice of the interfaces, for instance, imposing kinetic
energy constraints~\cite{ErpMoBol2003}.

The TIS method has been successfully applied to two realistic cases, the
folding of a polypeptide \cite{bolhuisPNAS} and hydration of ethylene
\cite{titusthesis}. In this last case the method was combined with quantum
ab-initio MD simulations. A recent variation of the TIS method for diffusive systems
exploits very efficiently the loss of long time scale correlation by using a
recursive reformulation of the crossing probability and the sampling of much
shorter paths~\cite{ErpMoBol2003_2}. These results show that TIS is capable of
studying rare event processes in complex systems efficiently and encourage even
more challenging applications, such as isomerization in clusters and crystal
nucleation, on which we plan to report in the future.


\bibliographystyle{prsty}
\bibliography{daniele}


\clearpage

%


\end{document}